\documentclass[10pt, conference, compsocconf]{IEEEtran}
\ifCLASSINFOpdf
\else
\fi
\usepackage{algorithm}
\usepackage[noend]{algpseudocode}
\usepackage{graphicx}
\usepackage{color}
\usepackage{float}
\floatstyle{plaintop}
\restylefloat{table}
\usepackage[table,xcdraw]{xcolor}
\usepackage{multirow}
\usepackage{subcaption}
\usepackage[export]{adjustbox}
\usepackage{cite}
\usepackage{amsmath}
\usepackage{graphicx}
\usepackage{bbm}
\usepackage[latin1]{inputenc}
\usepackage{multirow}
\usepackage{listings}
\usepackage{xcolor}
\graphicspath{{./Images/}}

\begin{document}
%
\title{Gene expression \\ for simulation of biological tissue}

\author{\IEEEauthorblockN{Sadyk Sayfullin, 
Manuel Mazzara,  Ruslan Mustafin and Victor Rivera}
\IEEEauthorblockA{
Innopolis University, Russia\\
Email: s.sayfullin@innopolis.ru, m.mazzara@innopolis.ru, r.mustafin@innopolis.ru and v.rivera@innopolis.ru}
\and\hspace{5cm}
\IEEEauthorblockN{Fedor Akhmetov}
\IEEEauthorblockA{
\hspace{5cm}National Research Nuclear University MEPhI, Russia\\
\hspace{5cm}Email: f.o.akhmetov@gmail.com}
}

%


\maketitle

\begin{abstract}
BioDynaMo is a biological processes simulator developed by an international community of researchers and software engineers working closely with neuroscientists. The authors have been working on gene expression, i.e. the process by which the heritable information in a gene - the sequence of DNA base pairs - is made into a functional gene product, such as protein or RNA. Typically, gene regulatory models employ either statistical or analytical approaches, being the former already well understood and broadly used. In this paper, we utilize analytical approaches representing the regulatory networks by means of differential equations, such as Euler and Runge-Kutta methods. The two solutions are implemented and have been submitted for inclusion in the BioDynaMo project and are compared for accuracy and performance.
\end{abstract}


%
\IEEEpeerreviewmaketitle

\vspace{-0.2cm}

\section{Introduction}
The mechanisms behind cells functioning have been progressively unveiled by the scientific community discovering the physical, biological and chemical principles on which they are operating. The problem that scientists are facing today is how to generate working hypotheses starting from the data collected over the years. There is still a limited number of software tools that can be effectively used for this kind of applications. Simulators have been proven to be a viable approach to interpret the results collected by experiments. 

BioDynaMo \cite{BreitwieserBMJK16} is one of these simulators developed by an international community of researchers and software engineers. 
This paper specifically focuses on the implementation of the biology module that allows simulating gene expression in the project developed by the researchers of BioDynaMo on top of the whole existing system\footnote{https://github.com/BioDynaMo/biodynamo}. 


\subsection{BioDynaMo} 
The BioDynaMo project \cite{BreitwieserBMJK16} is a simulator of biological processes designed to support the work of researchers in the biological field. Nowadays, a number of world-wide labs create their own software for running simulations. However, these solutions can often be utilized by specific labs and on specific tasks, without scaling and requiring significant resources.

BioDynaMo has been designed to address this problem in order to provide researchers with software support. The project is developing a new general platform for computer simulations of biological tissue dynamics, with brain development as a primary target. The platform should be executable on hybrid cloud computing systems, allowing for the efficient use of state-of-the-art computing technology. 

A set of different cellular behaviors is covered in this simulation such as cell division, cell growth, gene expression, chemical gradient and mechanical forces. Spatial locality of interaction is the main feature that makes this project run efficiently on highly parallelized cloud systems. This principle states that simulation objects reference to each other in the case when they are close\cite{denning2006locality}. It allows simulation space to be split up into fragments that do not require a large amount of communication between each other. Scale of simulation close to Cx3D\cite{zubler2009framework, cx3dp} upon which BioDynaMo was initially created\cite{bauer2017biodynamo, breitwieser2015}. However, with additional features, the project reaches simulation level of \cite{hoehme2010cell}.

The aim of the BioDynaMo project is to push the limits of this simulation type with both the highly efficient code and the extensive parallelization on relatively cheap cloud-based hardware\cite{bauer2017biodynamo}.

\subsection{Gene expression}
\begin{figure}[H]
\includegraphics[width=85mm]{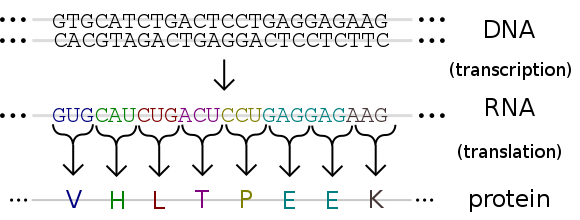}
\caption[qq]{Steps of gene expression. From Wikipedia
\footnotemark}
\label{fig:1}
\end{figure}

Gene expression is the process by which gene products, such as RNA and proteins, are produced. Sequences of DNA store heritable information that is used to produce gene product. Figure \ref{fig:1} depicts the main idea of this process: at first DNA is transcribed into RNA, which is subsequently translated into proteins \cite{misteli2001protein}. Proteins make many of the structures and all the enzymes in a cell or organism.
Several steps in the gene expression process might be modulated. This includes both the transcription and translation stages. Several biological processes controlled by gene expression and slight changes of specific proteins' concentration or links can underlie human diseases, population differences and the evolution of morphological novelties\cite{carroll2013dna}. In addition, types of cancer can be classified by tracking of gene expression\cite{golub1999molecular}.
\footnotetext{https://en.wikipedia.org/wiki/Gene\_expression\#/media/File:Genetic\_code.svg}
\section{Gene expression in simulations}
Computational models of gene expression can be used for various reasons. For instance, they enable the characterisation of the complex and dynamic relationship between genetic encoding and phenotypic expression \cite{karr2012whole}. Moreover, they can be used to generate \textit{in-silico} data to assess the performance and robustness of algorithms that infer gene-regulatory networks \cite{bansal2006inference}.

There are two main approaches in gene regulatory models: statistical and analytical. Statistical approaches have a high level of accuracy 
due to the number of
studies made on gene array data and exhaustively reviewed in the literature \cite{markowetz2007inferring, hecker2009gene}. In contrast,
the value of analytical approaches applied for modeling gene regulation is generally less appreciated, in particular in the field of DNA-sequence-based modeling. This paper shall examine the analytical approaches in order to investigate their potentiality and limitations. In general, analytical approaches concentrate on simulation of expression of a few genes and uses a mixture of different mathematical models in the implementation. In order to develop this approach, deep understanding of how parts of the system work individually is required, and hypotheses on how the composition of these parts behaves together. Analytical approaches can simulate terms relating to the binding of transcription factors and RNA polymerase to the DNA, cooperative and inhibitory interactions between transcription factors, mRNA and protein degradation, and mRNA translation rate \cite{ay2011mathematical}. 
\subsection{Thermodynamical model}
In thermodynamical models, gene production is regulated by bound activators and bound repressors. For a variety of mixture of these binding factors on regulatory regions, the thermodynamical model predicts the concentration of gene products. The main assumption in this model is that the level of production is proportional to bound activators and inversely proportional to bound regressors \cite{ay2011mathematical}.

\subsection{Boolean model}
Each gene product in this method has the property that represents its state "ON" and "OFF". To define the relationships between entities, logic functions  "and", "or" and "not" are used \cite{kauffman1969metabolic}. For example, if an expression of a gene is controlled by two products, the gene produces mRNA only if both products are ON in case of the function AND, the gene is transcribed if one of two products is on in case of OR function, and NOT means that gene is not transcribed if both are ON.
%
%
\subsection{Differential equation model}
Differential equations can be used to model gene regulation networks. In this model, interactions between entities produced by gene expression and concentration of them are defined by a set of differential equations. These equations depend on the variety of parameters, such as time, space, the concentration of other products such as mRNA and proteins and production and degradation rate of the particular entity \cite{chen1999modeling}.
\section{Implementation}
\subsection{Mathematical basis}
In BioDynaMo, a differential equation model has been implemented using Ordinary Differential Equations (ODE) which depend on a single variable i.e. time. The given task is shown by equation  (\ref{equation:task})
\begin{equation}
f(\textbf{p}(t),t) = \frac{d\textbf{p}(t)}{dt}
\label{equation:task}
\end{equation}
where $f(\textbf{p}(t),t)$ the given function of changing of protein concentration over time, $\textbf{p}(t)$ is the concentration of protein at time $t$. The idea is to track $\textbf{p}$ through time. Two methods have been implemented in order to solve this task:
\begin{itemize}
\item Runge-Kutta method
\item Euler method
\end{itemize}
These methods solve the Cauchy boundary value problem\cite{hadamard2014lectures} for differential equation $f(\textbf{p}(t),t)$ and given initial values $t_0$,  $\textbf{p}(t_0)=\textbf{p}_0$. Both methods calculate $\textbf{p}_n=\textbf{p}(t_n)$ step by step at $t_0, t_1, t_2\ldots$ .

\subsubsection{Runge-Kutta method}
In general, Runge-Kutta method for order $s$ is defined by equation (\ref{equation:RKGeneral})
\begin{equation}
\textbf{p}_n = \textbf{p}_{n-1}+h\sum_{i=1}^{s}b_iF_i, \ n=1,2,\ldots
\label{equation:RKGeneral}
\end{equation}
where $F_i$ is defined as
\begin{equation}
F_i = f(x_{n-1} +hc_i,\textbf{p}_{n-1} +h\sum_{j=1}^{s}a_{ij}F_j)
\end{equation}
where $a_{ij},c_i, b_i$ defined by tableau\cite{butcher2007runge} see Table \ref{table:tableauGeneral}:
\begin{table}[H]
\normalsize
\centering
\begin{tabular}{l| c c c c r}
 0&0&0&0&\ldots&0\\
 $c_2$&$a_{21}$&0&0&\ldots&0\\
 $c_2$&$a_{31}$&$a_{32}$&0&\ldots&0\\
 \vdots&\vdots&\vdots&\vdots&\vdots&\vdots\\
 $c_s$&$a_{s1}$&$a_{s2}$&$a_{s3}$&\ldots&0\\\hline
 &$b_1$&$b_2$&$b_3$&\ldots&$b_s$
\end{tabular}
\caption{Tableau for Runge-Kutta method of order $s$}
\label{table:tableauGeneral}
\end{table}
and $h=(t_{n}-t_{n-1})$ is time step of one iteration\\
In this work, we use Runge-Kutta of order 4. Thus,
it has an error rate $O(h^4)$ \cite{atkinson2008introduction}. For Runge-Kutta of order 4 tableau see Table \ref{table:tableauRK4}.
\begin{table}[H]
\normalsize
\centering
\begin{tabular}{l| c c c r}
0&0&0&0&0\\
$1/2$&$1/2$&0&0&0\\
$1/2$&0&$1/2$&0&0\\
1&0&0&1&0\\\hline
&$1/6$&$1/3$&$1/3$&$1/6$\\
\end{tabular}
\caption{Tableau for Runge-Kutta method of order 4}
\label{table:tableauRK4}
\end{table}
Thus, the Runge-Kutta method of order 4 is defined by equations (\ref{equation:RK4}) and (\ref{equation:Fs})
%
%
\begin{equation}
\textbf{p}_n = \textbf{p}_{n-1}+\frac{h}{6}(F_1+2F_2+2F_3+F_4)
\label{equation:RK4}
\end{equation}
\begin{equation}
\begin{split}
F_1 &= f(\textbf{p}_{n-1}, t_{n-1})
\\ 
F_2 &= f(\textbf{p}_{n-1} + \frac{h}{2}F_1, t_{n-1} + \frac{h}{2})
\\
F_3 &= f(\textbf{p}_{n-1} + \frac{h}{2}F_2, t_{n-1} + \frac{h}{2})
\\
F_4 &= f(\textbf{p}_{n-1} + hF_4, t_{n-1} + h)
\end{split}
\label{equation:Fs}
\end{equation}
\begin{algorithm}[H]
\label{algorithm:Runge-Kutta}
\normalsize
\caption{Runge-Kutta method}\label{alg:search}
\begin{algorithmic}[1]
\Procedure{search}{$c\_t$, $t\_s$, $p$}\vspace{0.5mm}
\State $F1$=$Equation(c\_t,\ p)$\vspace{0.5mm}
\State $F2$=$Equation(c\_t+t\_s/2,\ p+t\_s*F1/2)$\vspace{0.5mm}
\State $F3$=$Equation(c\_t+t\_s/2,\ p+t\_s*F2/2)$\vspace{0.5mm}
\State $F4$=$Equation(c\_t+t\_s,\ p+t\_s*F3)$\vspace{0.5mm}
\State $p=p+t\_s*(F1+2*F2+2*F3+F4)/6$\vspace{0.5mm}
\State $c\_t = c\_t + t\_s$
\EndProcedure
\end{algorithmic}
\end{algorithm}
Where $c\_t$ is current time $t_n$, $t\_s$ time step $h$, $p$ is protein concentration $\textbf{p}$,  $Equation$ is the given differential equation $f(\textbf{p}(t),t)$ and $F1$ - $F4$ parameters from equation (\ref{equation:Fs})

The Runge-Kutta method has high accuracy but the performance of this method is lower in comparison with the Euler method due to the amount of operations it carries out: 4 calls to $Equation$ against 1 call in Euler method.

\subsubsection{Euler method}
The Euler method is the Runge-Kutta method of order $s=1$. For Euler method tableau see Table \ref{table:tableauEuler}:
\begin{table}[H]
\normalsize
\centering
\begin{tabular}{l| r}
0&0\\\hline
&1\\
\end{tabular}
\caption{Tableau for Euler method (Runge-Kutta method of order 1)}
\label{table:tableauEuler}
\end{table}
The Euler method \cite{butcher2007runge} is defined  as equation (\ref{equation:Euler})
\begin{equation}
\textbf{p}_n = \textbf{p}_{n-1}+h\cdot f(\textbf{p}_{n-1},t_{n-1}), \ n=1,2,\ldots
\label{equation:Euler}
\end{equation}
The Euler method at each iteration makes only one operation and once calls for $f(\textbf{p}(t),t)$. This allows the calculations to be faster than the previous method. However, the error rate of this method is bigger $O(h)$\cite{atkinson2008introduction}.
\begin{algorithm}[H]
\label{algorithm:Euler}
\normalsize
\caption{Euler method}\label{alg:search}
\begin{algorithmic}[1]
\Procedure{compute}{$current\_time$, $time\_step$, $protein$}\vspace{0.5mm}
\State $p$=$p+t\_s$*$Equation(c\_t,p)$\vspace{0.5mm}
\State $c\_t = c\_t + t\_s$\vspace{0.5mm}
\EndProcedure
\end{algorithmic}
\end{algorithm}

\subsubsection{Comparison}
Both methods are used to solve the same task. However, they differ in accuracy and performance. The error rate of Runge-Kutta method $O(h^4)$ and $O(h)$ for Euler method. At the same time, Runge-Kutta has lower performance
due to the number of
operations on each iteration.

\subsection{Biology module implementation}
Simulation objects and modules for different types of simulations are the main parts of BioDynaMo project.
On the stage of initialization of simulation objects, a unique copy of module binds with each object. This allows to store and work with additional information about particular cell without changes of simulation object's class.

\begin{figure}[H]
\centering
\includegraphics[width=80mm]{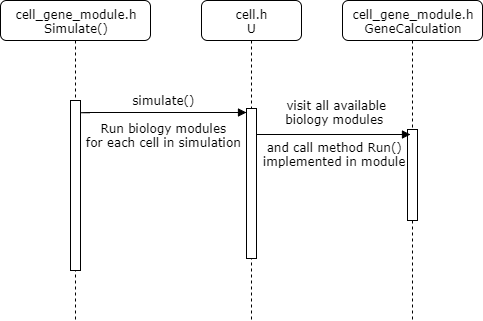}
\caption{Sequence diagram of main actions of one iteration of simulation}
\label{fig:SeqD1}
\end{figure}

Figure \ref{fig:SeqD1} depicts simplified sequence diagram of one iteration for biology module \textit{GeneCalculation}. It starts from calling of method \textit{Simulate} from \textit{Scheduler}. Then \textit{Scheduler} for each cell in simulation looks for bound with it modules. As the final step, method \textit{Run()} in every found module is called. 


Listing \ref{lst:GeneCalc} shows the implementation of biology module \textit{GeneCalculation}. This module simulates expression of genes and contains all required additional variables for tracking of the concentration of proteins. Thus, can work with any type of simulation object.
It has the implementation of Euler and Runge-Kutta numerical methods for solving ODE which are essential in the picked model. Both methods implemented inside the body of method \textit{Run()}. The user determines which method is picked in particular simulation  through variable \textit{DE\_solve\_method} from structure \textit{Param}.
Variable \textit{substances\_} stores current value for each simulating protein. Lambda function \textit{functions} from structure \textit{Param} is defined by user and it is required for this function to return vector of values calculates all given functions $f(\textbf{p}(t),t)$.

Structure \textit{Param} contains variables that determine parameters of the simulation. To this structure were added variables that are essential for gene expression simulation such as \textit{DE\_solve\_method} and \textit{functions}.




\begin{lstlisting}[caption={Implementation of biology module \textit{GeneCalculation}},label={lst:GeneCalc}]
struct GeneCalculation : public BaseBiologyModule {
    double time_step = Param::simulation_time_step;
    vector<array<double, Param::protein_amount>>
    	substances_;

    GeneCalculation() : 
   	 BaseBiologyModule(gAllBmEvents), 
   	 substances_(init_vals){}

    template <typename T>
    void Run(T* cell) {
      if (Param::DE_solve_method == "Euler"){
        vector<double> update_value = 
         Param::functions(Param::step_global_ *
         time_step, substances_);
        for (int i = 0;
        	i < Param::protein_amount; i++){
          substances_[0][i] += 
           update_value[i] * time_step;
        }
      }
      else if (Param::DE_solve_method == "RK4"){
        vector<double> k1 = 
         Param::functions(Param::step_global_ *
         time_step, substances_);
         
        for (int i = 0;
         i < Param::protein_amount; i++)
          substances_[0][i] +=time_step*k1[i]/2.0f;
          
        vector<double> k2 = 
         Param::functions(Param::step_global_ *
         time_step + time_step/2.0f, substances_);
         
        for (int i = 0;
         i < Param::protein_amount; i++)
          substances_[0][i] += time_step*k2[i]/2.0f
          - time_step*k1[i]/2.0f;
            
        vector<double> k3 = 
         Param::functions(Param::step_global_ * 
         time_step + time_step/2.0f, substances_);
        
        for (int i = 0;
         i < Param::protein_amount; i++)
          substances_[0][i] += time_step*k3[i] -
          time_step*k2[i]/2.0f;
          
        vector<double> k4 = 
         Param::functions(Param::step_global_ * 
         time_step + time_step, substances_);
        
        for (int i = 0;
        i < Param::protein_amount; i++){
          substances_[0][i] += time_step*(k1[i] +
           2*k2[i] + 2*k3[i] + k4[i])/6.0f;
        }
      }
    }
    ClassDefNV(GeneCalculation, 1);
  };
\end{lstlisting}

Both algorithms for Euler and Runge-Kutta methods were implemented in C++ and experiments were run on Intel\textsuperscript{\textregistered} Core\texttrademark i5-6200U CPU @ 2.30GHz $\times$ 4. \\

\vspace{-0.7cm}
\section{Results}
\begin{figure*}[tb]
\begin{subfigure}[tb]{0.5\textwidth}
\includegraphics[width=90mm]{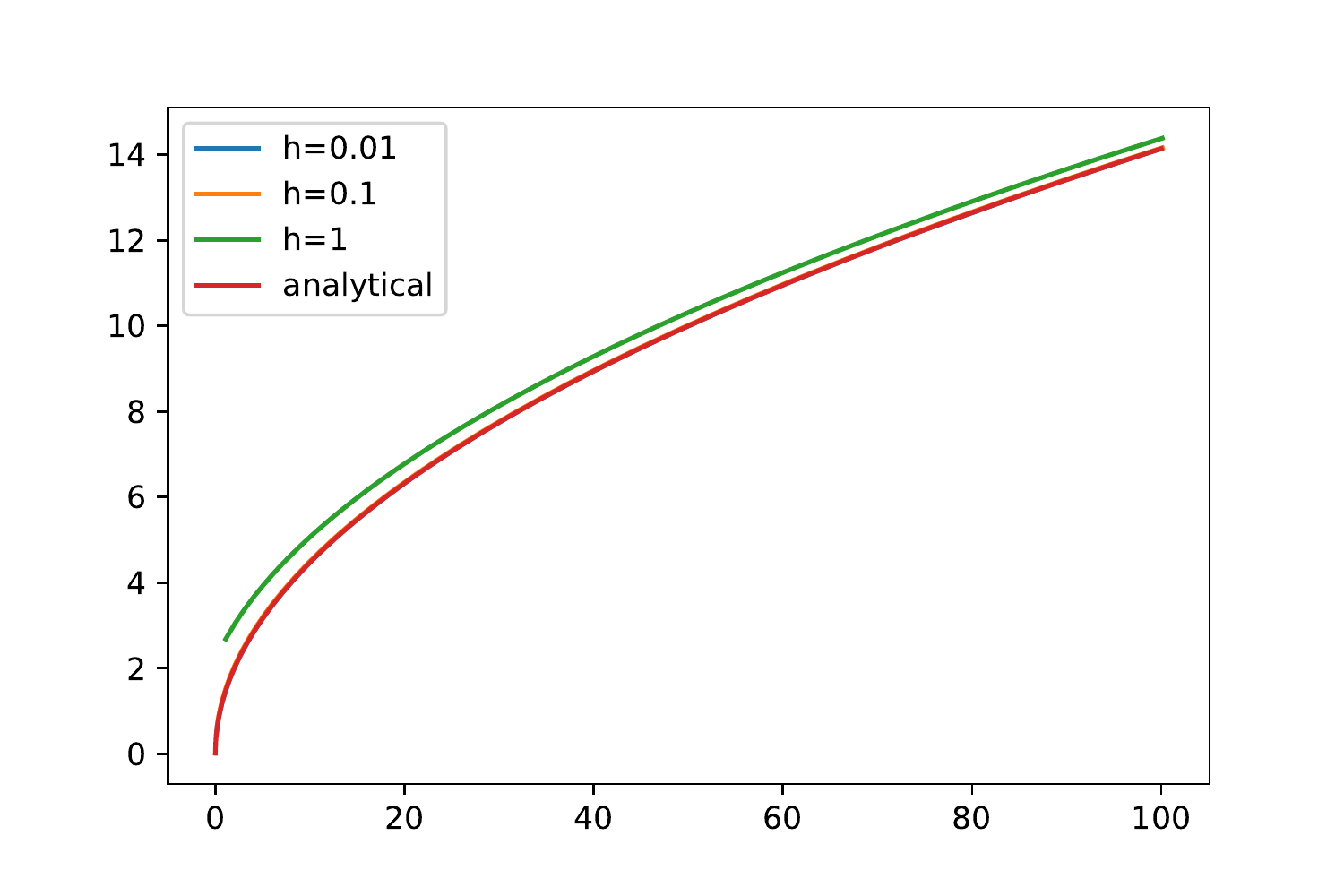}
\centering
\caption{}
\label{fig:2}
\end{subfigure}
\begin{subfigure}[tb]{.5\textwidth}
\includegraphics[width=90mm]{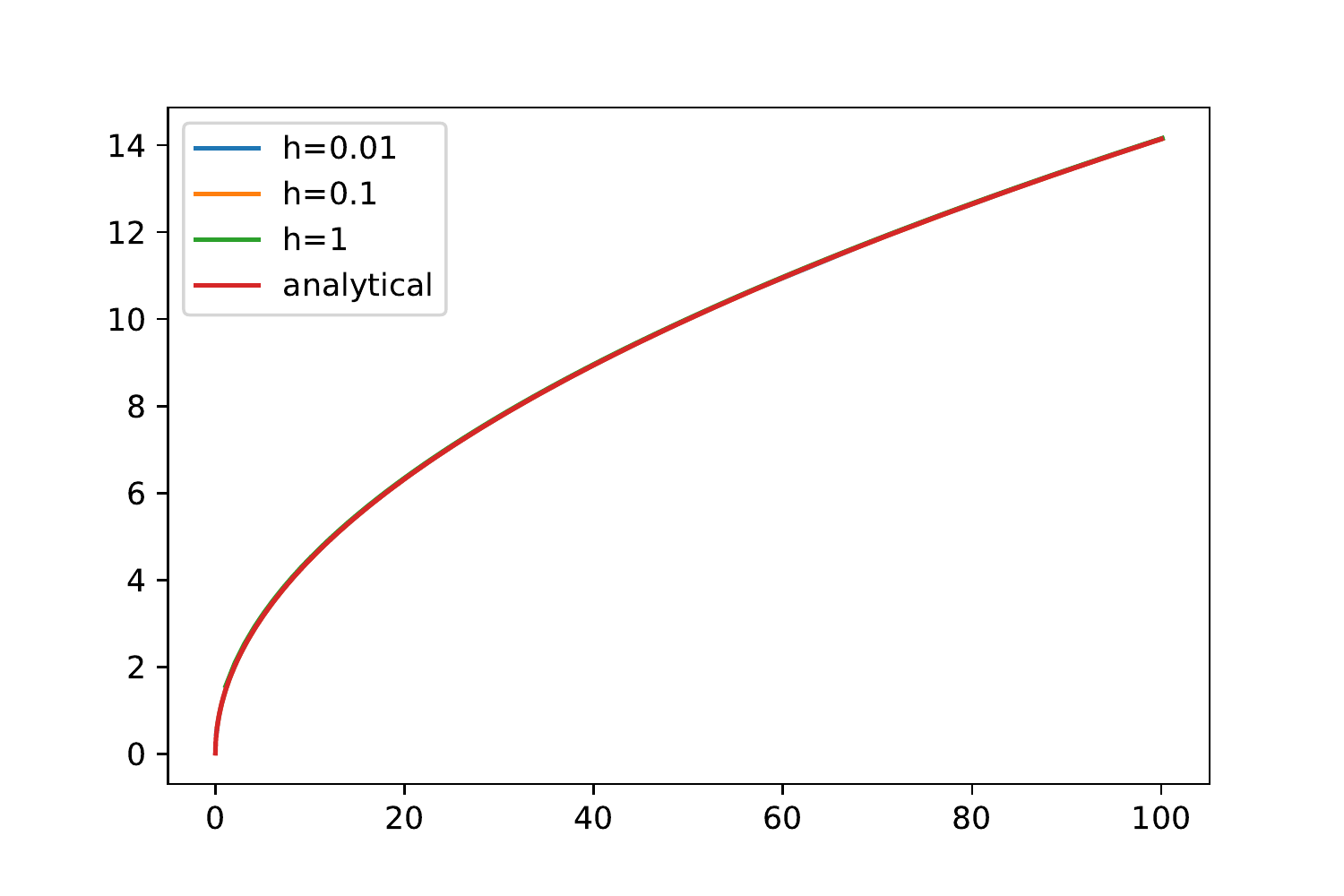}
\caption{}
\label{fig:3}
\end{subfigure}
\caption{The plot for function (\ref{equation:1/tInit}) made by Euler (a), Runge-Kutta (b) method with different $h$ parameters and plot for analytical solution (\ref{equation:1/tAnSol})}
\label{fig:equat1}
\end{figure*}

\begin{figure}[h]
\begin{subfigure}{0.5\textwidth}
\includegraphics[width=90mm]{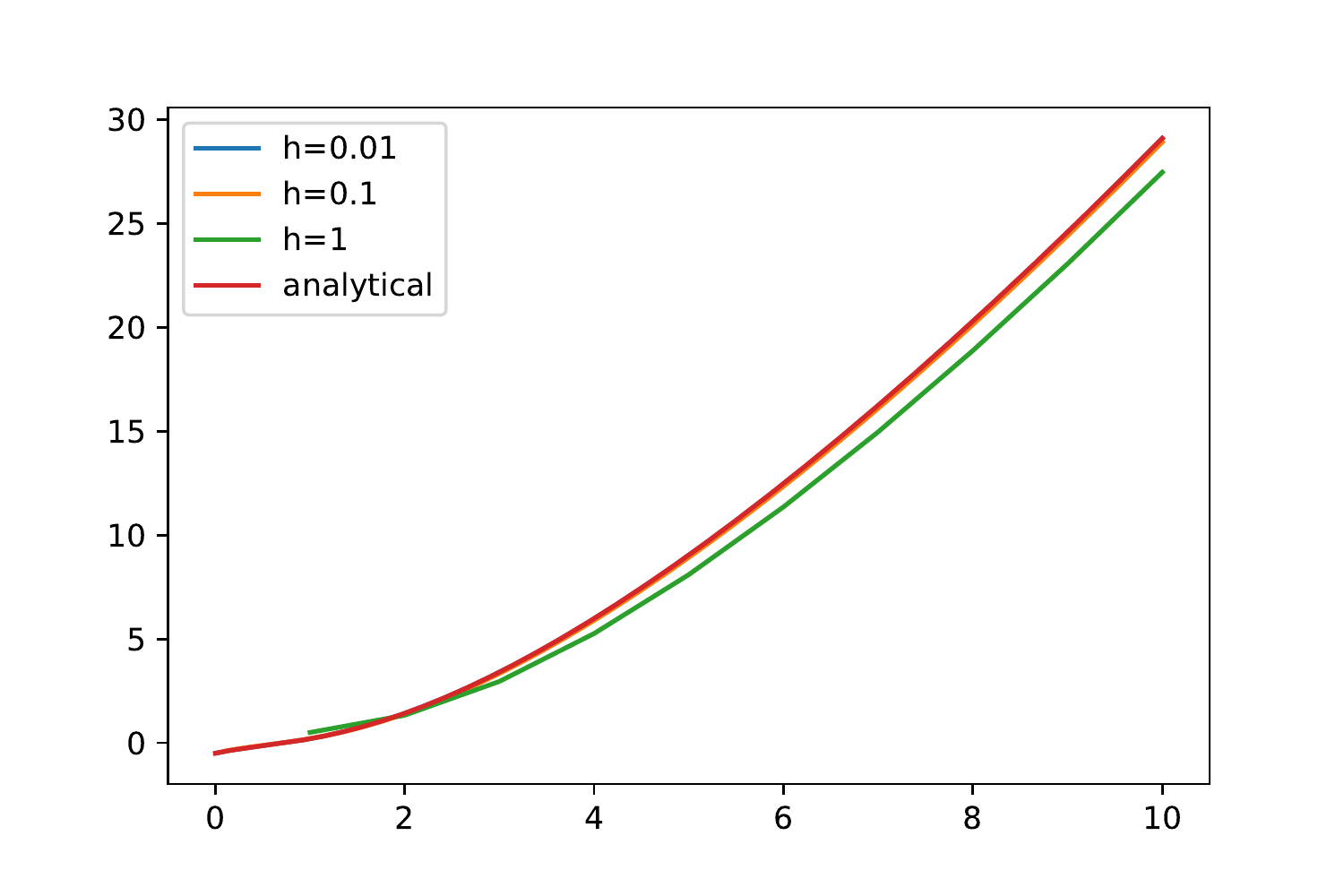}
\centering
\caption{}
\label{fig:4}
\end{subfigure}\\
\begin{subfigure}{.5\textwidth}
\includegraphics[width=90mm]{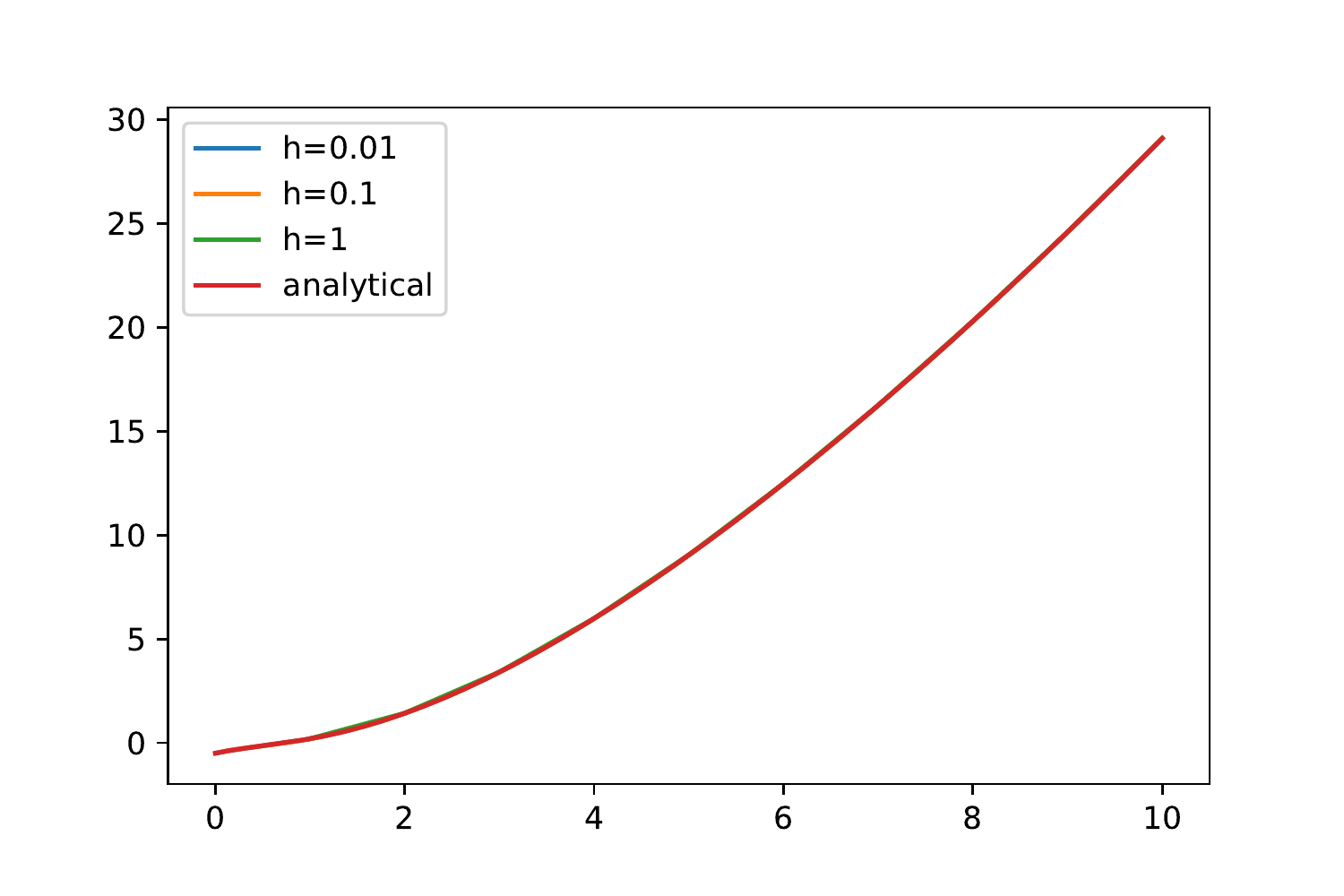}
\caption{}
\label{fig:5}
\end{subfigure}\\
\caption{Plots for function (\ref{equation:logInit}) made by Euler (a), Runge-Kutta (b) methods with different $h$ parameters and plot for analytical solution (\ref{equation:logAnSol})}
\label{fig:equat2}
\end{figure}

\begin{figure}[h]
\begin{subfigure}{0.5\textwidth}
\includegraphics[width=90mm]{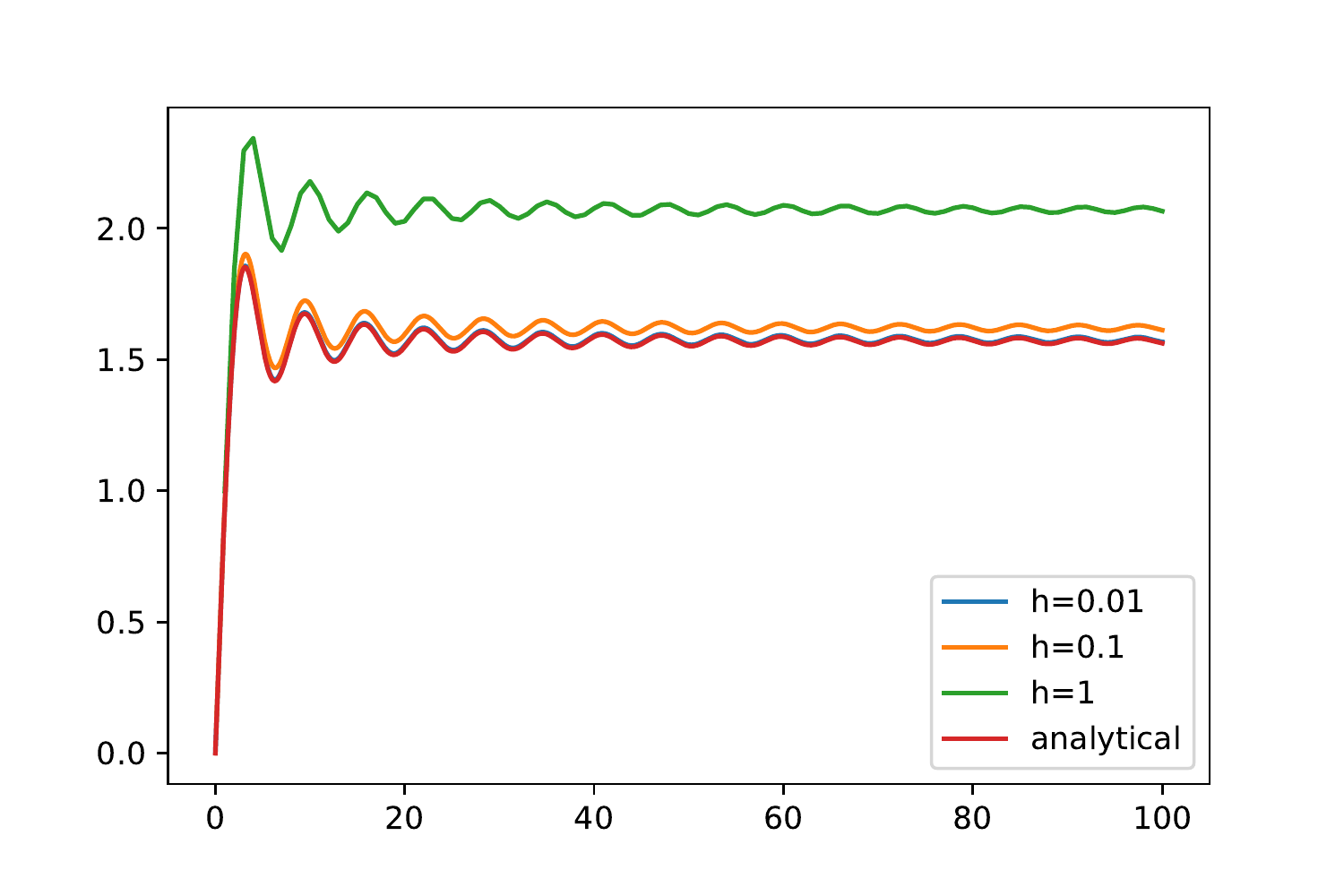}
\centering
\caption{}
\label{fig:6}
\end{subfigure}\\
\begin{subfigure}{.5\textwidth}
\includegraphics[width=90mm]{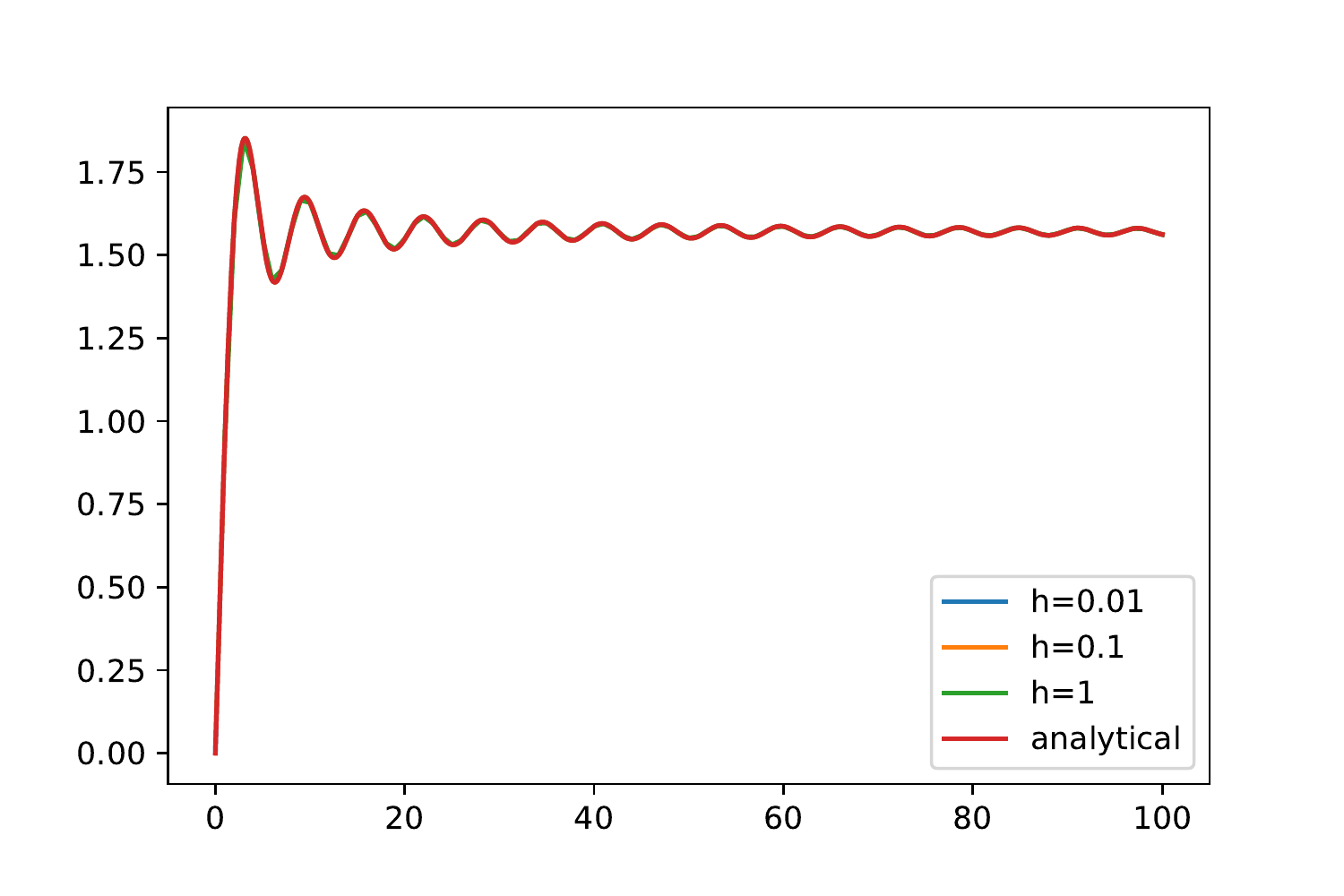}
\caption{}
\label{fig:7}
\end{subfigure}\\
\caption{The plot for function (\ref{equation:sinInit}) made by Euler (a), Runge-Kutta (b) methods with different $h$ parameters and plot for analytical solution (\ref{equation:sinAnSol})}
\label{fig:equat3}
\end{figure}

Figures \ref{fig:equat1}, \ref{fig:equat2}, \ref{fig:equat3} and Table \ref{table:2} present the results of the solution to three Cauchy problems for equations (\ref{equation:1/tInit}), (\ref{equation:logInit}) and (\ref{equation:sinInit}). Solutions are presented using the Euler and Runge-Kutta numerical methods with different parameter $h$. Furthermore, they present the analytical solution for the same equations in order to evaluate the accuracy of the methods. Each cell in Table \ref{table:2} displays the difference between analytical and numerical solutions.

Figure \ref{fig:equat1} depicts plots for the Cauchy problem (\ref{equation:1/tInit})
\begin{equation}
\label{equation:1/tInit}
	\begin{cases}
		f(\textbf{p}(t),t) = \frac{1}{t}\\
        \textbf{p}_{0}=\textbf{p}(0)=0
    \end{cases}
\end{equation}
The analytical solution for this task is depicted by equation (\ref{equation:1/tAnSol}):
\begin{equation}
\label{equation:1/tAnSol}
\textbf{p}(t) = \sqrt[]{2t}
\end{equation}
Figure \ref{fig:2} shows the solution by the Euler method. It is observed that for $h = 0.1$, $h = 0.01$ and the analytical solution the lines overlap. The plot with $h = 1$ does not give correct values from the beginning.




\begin{table*}[ht]\normalsize
\begin{tabular*}{1\textwidth}{@{\extracolsep{\fill}}| l  l | c | c | c | c| c | c | r | }
\hline
\multicolumn{1}{|l}{}&\multicolumn{1}{|l|}{}&\multicolumn{3}{c|}{Euler}&\multicolumn{3}{c|}{Runge-Kutta}&\multicolumn{1}{r|}{}\\\cline{2-9}
\multirow{5}{*}{Equation \ref{equation:1/tInit}} &\multicolumn{1}{|l|}{time}&$h=0.01$&$h=0.1$&$h=1$&$h=0.01$&$h=0.1$&$h=1$&analytical\\\cline{2-9}
&\multicolumn{1}{|l|}{1}&0.004081&0.045621&1.20004&3.026e-7&4.03e-5&0.07955&1.48324\\\cline{2-9}
&\multicolumn{1}{|l|}{10}&0.002578&0.027378&0.591468&-1.011e-6&8.989e-6&0.02693&4.49444\\\cline{2-9}
&\multicolumn{1}{|l|}{100}&0.001195&0.012795&0.235295&-4.925e-6&-4.925e-6&0.00859&14.14920\\\cline{2-9}
&\multicolumn{1}{|l|}{500}&0.000661&0.006961&0.118461&-3.872e-5&-3.872e-5&0.00386&31.62594\\\cline{2-9}
&\multicolumn{1}{|l|}{1000}&0.000504&0.005304&0.087704&4.438e-6&4.438e-6&0.00270&44.72359\\\hline
\multicolumn{1}{|l}{}&\multicolumn{1}{|l|}{}&\multicolumn{3}{c|}{Euler}&\multicolumn{3}{c|}{Runge-Kutta}&\multicolumn{1}{r|}{}\\\cline{2-9}
\multirow{5}{*}{Equation \ref{equation:logInit}} &\multicolumn{1}{|l|}{time}&$h=0.01$&$h=0.1$&$h=1$&$h=0.01$&$h=0.1$&$h=1$&analytical\\\cline{2-9}
&\multicolumn{1}{|l|}{1}&0.00088&0.010849&0.303724&1.342e-7&1.342e-7&0.002486&0.196275\\\cline{2-9}
&\multicolumn{1}{|l|}{10}&-0.018060&-0.178960&-1.636460&3.948e-5&3.948e-5&0.003139&29.093460\\\cline{2-9}
&\multicolumn{1}{|l|}{100}&-0.040630&-0.408630&-3.948630&0.000369&0.000369&0.003369&724.165630\\\cline{2-9}
&\multicolumn{1}{|l|}{500}&-0.057691&-0.567691&-5.557691&0.002309&0.002309&0.002309&5217.74769\\\cline{2-9}
&\multicolumn{1}{|l|}{1000}&-0.051150&0.651150&-6.251150&0.048849&0.048849&0.048849&11818.6511\\\hline
\multicolumn{1}{|l}{}&\multicolumn{1}{|l|}{}&\multicolumn{3}{c|}{Euler}&\multicolumn{3}{c|}{Runge-Kutta}&\multicolumn{1}{r|}{}\\\cline{2-9}
\multirow{5}{*}{Equation \ref{equation:sinInit}} 
&\multicolumn{1}{|l|}{time}&$h=0.01$&$h=0.1$&$h=1$&$h=0.01$&$h=0.1$&$h=1$&analytical\\\cline{2-9}
&\multicolumn{1}{|l|}{1}&0.000789&0.007675&0.053916&-7.037e-8&-7.037e-8&6.293e-5&0.946083\\\cline{2-9}
&\multicolumn{1}{|l|}{10}&2.406e-6&2.406e-6&2.241e-5&0.005272&0.052652&0.520572&1.658347\\\cline{2-9}
&\multicolumn{1}{|l|}{100}&4.533e-6&4.533e-6&-5.467e-6&0.005024&0.050264&0.503264&1.562225\\\cline{2-9}
&\multicolumn{1}{|l|}{500}&0.005004&0.050044&0.500314&4.118e-6&4.118e-6&4.118e-6&1.572565\\\cline{2-9}
&\multicolumn{1}{|l|}{1000}&0.004997&0.049957&0.499637&-3.122e-6&-3.122e-6&-3.122e-6&1.570233\\\hline
\end{tabular*}
\caption{Accuracy of Euler and Runge-Kutta methods}
\label{table:2}
\end{table*}

Figure \ref{fig:3} shows the solution by the Runge-Kutta method. It is observed that lines for all $h$ and the analytical solution overlap.

Figure \ref{fig:equat2} presents plots for the Cauchy problem \ref{equation:logInit}:
\vspace{0.5cm}
\begin{equation}
\label{equation:logInit}
\begin{cases}
		f(\textbf{p}(t),t) = log(1+t^2)+\exp^{-2t}\\
        \textbf{p}_{0}=\textbf{p}(0)=-\frac{1}{2}
   \end{cases}
   \vspace{0.5cm}
\end{equation}
The analytical solution for this task is equation (\ref{equation:logAnSol}):
\begin{equation}
\label{equation:logAnSol}
\textbf{p}(t) = t\cdot log(t^2+1)-2t+2arctan(t)-\frac{1}{2}\exp^{-2t}
\end{equation}
Figure \ref{fig:4} depicts the solution by the Euler method. It is observed that lines for  $h=0.1$, $h=0.01$ and the analytical solution overlap. The line for the Euler method with parameter $h = 1$ diverges as the values increase.
Figure \ref{fig:5} depicts the solution by the Runge-Kutta method. Lines for all $h$ and the analytical solution overlap. 

Figure \ref{fig:equat3} shows plots for the Cauchy problem \ref{equation:sinInit}
\begin{equation}
\label{equation:sinInit}
\begin{cases}
		f(\textbf{p}(t),t) = \frac{sin(t)}{t}\\
        \textbf{p}_{0}=\textbf{p}(0)=0
   \end{cases}
\end{equation}
The analytical solution for this task is:
\begin{equation}
\label{equation:sinAnSol}
\textbf{p}(t) = Si(t)
\end{equation}
$Si(t)$ is the sine integral function defines as
\begin{equation}
\label{equation:Si}
Si(t) = \int_{0}^t\frac{sin(z)}{z}dz
\end{equation}
Figure \ref{fig:6} presents the solution by the Euler method. It is perceived that the line for $h = 0.01$ and the analytical solution overlap. Although, the lines for higher $h$ diverges from the analytical solution, they repeat the shape. However, the values obtained with $h = 1$ are unreliable. 

Figure \ref{fig:7} depicts the solution by the Runge-Kutta method. Lines for all $h$ and analytical solution distinctly overlap. 

\section{Discussion}
The BioDynaMo project has a set of classes to deal with different types of simulation: biological, physical and diffusion. The biological simulation consist of different modules, for example: \textit{GrowthDivide}, \textit{GrowthModule}, \textit{Chemotaxis}. Each cell can carry an individual set of biology modules.

As an extension to the project, a new biology module \textit{GeneExpression} has been added. This module stores protein values for each cell. Furthermore, 
this module allows custom laws definition
in the form of ODEs that define protein concentration. To solve sets of ODEs, a user may select from two available options depending on their current needs: the Euler method for higher performance and the Runge-Kutta method for accuracy. However, in the case when the simulation time step is small, in order to speed up the simulation, the Euler method can be used without a notable loss of accuracy. In addition, the user specifies functions $f(\textbf{p}(t),t)$ by which changes in concentration of proteins is calculated and initial values for proteins.
%
%
%
%
%
%


\section{Related works}
Hucka et al. \cite{sbml} presented a markup language to describe biochemical network models named SBML. One of their primary goals was to create a common format that can be used across simulators. Several implementation exist for this specification \cite{libroadrunner}, \cite{jsbml}.

The work presented in \cite{ramis2009multi} uses differential equations in order to model changes in the concentration of proteins as well. Although, this work uses gene expression in a specific task, i.e. studying the influence of different protein pathways in the spreading of cancer cells.

The article\cite{hecker2009gene} presents the reconstruction of gene regulatory networks (GRNs) from experimental data of gene expression through computational methods. The study describes models for reverse engineering GRNs from gene expression data. Along with Boolean and Bayesian models, the Differential equation model is described.

The research \cite{kim2004dynamic} introduces a new statistical gene network estimation method based on the dynamic Bayesian network and nonparametric regression model with advantages over Bayesian and Boolean networks. For example, it can detect nonlinear dependencies. This article can be taken into consideration in further work if the statistical approach will be implemented in BioDynaMo.

\section{Conclusion}
BioDynaMo is a simulator of biological processes developed by an international community of researchers and software engineers working in close synergy in order to implement the  requirements coming from neuroscientists. Bridging the gap between the neuroscience perspective and the technical software engineering perspective is one of the tasks of our team within the project. Collecting requirements from specialists and implementing them is a non-trivial task. We advocate the necessity of developing a requirements engineering framework offering different syntaxes to represent the same concept (text, graphical, mathematical) in order to facilitate the communication between the stakeholders without forcing them to change their current habits. For example, a comprehensive modeling framework as discussed in \cite{YanCZM07} may solve the problem.

Among the requirements to be understood and implemented, our research team focused  on the modeling of gene expression in the simulation, which is discussed in detail in this paper. We discussed the entire process of modeling and development presenting the problem, the possible solving techniques, providing multiple implementations, and comparing them under qualitative and qualitative aspects. 

Simulation of biologic dynamics has several applications, some direct and some indirect. Computer simulations of biological tissue dynamics can serve the purpose of understanding diseases and dysfunction at the operational and sub-operational level as well as being the effective substitute for drug testing that does not involve living beings. At the same time, potential application scenarios are also foreseeable in cognitive architecture \cite{Vallverdu16}, where simulations may provide insight into and, in general, for the development of smart systems, including smart houses \cite{Nalin2016, Salikhov2016a, Salikhov2016b} and smart automotive systems \cite{Gmehlich13}.

Finally, the entire project may benefit in terms of scalability if re-engineered in order to deploy in a flexible and continuous fashion, for example following the microservice paradigm \cite{Dragoni2017, DragoniLLMMS17}.
\bibliographystyle{IEEEtran}
\bibliography{IEEEabrv,mainbib}

\end{document}